\documentstyle[aps,prl,epsf]{revtex}
\newcommand {\ebar}{\hbox{$E$\kern-0.5em\lower-0.1ex\hbox{/}}} 
\begin{document}

\setcounter{page}{1}

\title{
\begin{flushright}
FERMILAB-PUB-00/064-E 
\end{flushright}
Limits on Gravitino Production and New Processes  with  
Large Missing Transverse Energy in $p\bar p$ Collisions at $\sqrt{s}=1.8$ TeV}
\maketitle

\font\eightit=cmti8
\def\r#1{\ignorespaces $^{#1}$}
\hfilneg
\begin{sloppypar}
\noindent
T.~Affolder,\r {21} H.~Akimoto,\r {43}
A.~Akopian,\r {36} M.~G.~Albrow,\r {10} P.~Amaral,\r 7 S.~R.~Amendolia,\r {32} 
D.~Amidei,\r {24} K.~Anikeev,\r {22} J.~Antos,\r 1 
G.~Apollinari,\r {36}
% ...
\end{sloppypar}
\vskip .026in
\begin{center}
(CDF Collaboration)
\end{center}

\vskip .026in
\begin{center}
\r 1  {\eightit Institute of Physics, Academia Sinica, Taipei, Taiwan 11529, 
Republic of China} \\
\r 2  {\eightit Argonne National Laboratory, Argonne, Illinois 60439} \\
\r 3  {\eightit Istituto Nazionale di Fisica Nucleare, University of Bologna,
I-40127 Bologna, Italy} \\
\r 4  {\eightit Brandeis University, Waltham, Massachusetts 02254} \\
\r 5  {\eightit University of California at Los Angeles, Los 
Angeles, California  90024} \\  
\r 6  {\eightit Instituto de Fisica de Cantabria, University of Cantabria, 
39005 Santander, Spain} \\
\r 7  {\eightit Enrico Fermi Institute, University of Chicago, Chicago, 
Illinois 60637} \\
\r 8  {\eightit Joint Institute for Nuclear Research, RU-141980 Dubna, Russia}
\\
\r 9  {\eightit Duke University, Durham, North Carolina  27708} \\
\r {10}  {\eightit Fermi National Accelerator Laboratory, Batavia, Illinois 
60510} \\
\r {11} {\eightit University of Florida, Gainesville, Florida  32611} \\
\r {12} {\eightit Laboratori Nazionali di Frascati, Istituto Nazionale di Fisica
               Nucleare, I-00044 Frascati, Italy} \\
\r {13} {\eightit University of Geneva, CH-1211 Geneva 4, Switzerland} \\
\r {14} {\eightit Harvard University, Cambridge, Massachusetts 02138} \\
\r {15} {\eightit Hiroshima University, Higashi-Hiroshima 724, Japan} \\
\r {16} {\eightit University of Illinois, Urbana, Illinois 61801} \\
\r {17} {\eightit The Johns Hopkins University, Baltimore, Maryland 21218} \\
\r {18} {\eightit Institut f\"{u}r Experimentelle Kernphysik, 
Universit\"{a}t Karlsruhe, 76128 Karlsruhe, Germany} \\
\r {19} {\eightit Korean Hadron Collider Laboratory: Kyungpook National
University, Taegu 702-701; Seoul National University, Seoul 151-742; and
SungKyunKwan University, Suwon 440-746; Korea} \\
\r {20} {\eightit High Energy Accelerator Research Organization (KEK), Tsukuba, 
Ibaraki 305, Japan} \\
\r {21} {\eightit Ernest Orlando Lawrence Berkeley National Laboratory, 
Berkeley, California 94720} \\
\r {22} {\eightit Massachusetts Institute of Technology, Cambridge,
Massachusetts  02139} \\   
\r {23} {\eightit Institute of Particle Physics: McGill University, Montreal 
H3A 2T8; and University of Toronto, Toronto M5S 1A7; Canada} \\
\r {24} {\eightit University of Michigan, Ann Arbor, Michigan 48109} \\
\r {25} {\eightit Michigan State University, East Lansing, Michigan  48824} \\
\r {26} {\eightit University of New Mexico, Albuquerque, New Mexico 87131} \\
\r {27} {\eightit The Ohio State University, Columbus, Ohio  43210} \\
\r {28} {\eightit Osaka City University, Osaka 588, Japan} \\
\r {29} {\eightit University of Oxford, Oxford OX1 3RH, United Kingdom} \\
\r {30} {\eightit Universita di Padova, Istituto Nazionale di Fisica 
          Nucleare, Sezione di Padova, I-35131 Padova, Italy} \\
\r {31} {\eightit University of Pennsylvania, Philadelphia, 
        Pennsylvania 19104} \\   
\r {32} {\eightit Istituto Nazionale di Fisica Nucleare, University and Scuola
               Normale Superiore of Pisa, I-56100 Pisa, Italy} \\
\r {33} {\eightit University of Pittsburgh, Pittsburgh, Pennsylvania 15260} \\
\r {34} {\eightit Purdue University, West Lafayette, Indiana 47907} \\
\r {35} {\eightit University of Rochester, Rochester, New York 14627} \\
\r {36} {\eightit Rockefeller University, New York, New York 10021} \\
\r {37} {\eightit Rutgers University, Piscataway, New Jersey 08855} \\
\r {38} {\eightit Texas A\&M University, College Station, Texas 77843} \\
\r {39} {\eightit Texas Tech University, Lubbock, Texas 79409} \\
\r {40} {\eightit Istituto Nazionale di Fisica Nucleare, University of Trieste/
Udine, Italy} \\
\r {41} {\eightit University of Tsukuba, Tsukuba, Ibaraki 305, Japan} \\
\r {42} {\eightit Tufts University, Medford, Massachusetts 02155} \\
\r {43} {\eightit Waseda University, Tokyo 169, Japan} \\
\r {44} {\eightit University of Wisconsin, Madison, Wisconsin 53706} \\
\r {45} {\eightit Yale University, New Haven, Connecticut 06520} \\
\end{center}

\vskip .025in

\begin{abstract}
Events collected by the Collider Detector at Fermilab (CDF) 
with an energetic jet plus large missing transverse energy  
can be used to search for  physics beyond the Standard Model.
We see no deviations from the expected backgrounds and set upper limits on
the production of  new processes.
We consider in addition  the production of light 
gravitinos within the framework of the Gauge Mediated Supersymmetry Breaking 
models and set a limit at 95\% confidence level on the breaking scale 
$\sqrt{F}\ge 217$ GeV, which excludes gravitino masses smaller
 than $1.1\times 10^{-5}$ eV/c$^2$. 

\end{abstract}

\centerline{PACS numbers: 14.80.-j, 13.85.Rm, 13.87.Ce}

\newpage

\vspace{.3cm}
In $p\bar p$ collisions undetectable particles manifest themselves
 as missing transverse
energy, $\ebar_T$.
Events characterized by  large amounts of 
$\ebar_T$ are  interesting for searches of physics beyond 
the Standard Model.
Supersymmetry, for instance, relates each
 bosonic/fermionic Standard Model particle to a fermionic/bosonic
superpartner, providing a solution to the hierarchy problem\cite{susy}.
In supersymmetric models with Gauge Mediated Supersymmetry Breaking (GMSB), the
goldstino, a massless and neutral spin-$\frac{1}{2}$ particle, is introduced.
When gravitation is added and supersymmetry is realized locally the gauge
particle, graviton, has a spin-$\frac{3}{2}$ partner, the gravitino 
($\tilde G$), which
acquires a mass, $m_{\tilde G}$,  while the goldstino is absorbed \cite{gmsb}.
\par
At the Tevatron, gravitinos can be produced in pairs in association
either  with radiation 
jets according to the processes 
$q\bar q\to \tilde G\tilde Gg$, $qg\to \tilde G\tilde Gq$, $\bar qg\to \tilde
G\tilde G\bar q$ and $gg\to \tilde G\tilde Gg$, or with a photon following
$q\bar q\to {\tilde G}{\tilde G}\gamma$.
In the scenario in which all other supersymmetric particles are heavy, 
the main parameter upon which these processes depend is the
supersymmetry-breaking scale $\sqrt{F}$\cite{FF}
and the cross sections vary 
as $1/m_{\tilde G}^4$\cite{brig}.  
If supersymmetry is present, and the gravitino is very light 
($m_{\tilde G}\ll 10^{-4}$ eV/c$^2$), 
it can  be  seen at
the Tevatron by looking at final states which include gravitinos and ordinary
particles only\cite{brig}. In this case the lightest supersymmetric particle 
is the gravitino which escapes undetected
manifesting itself as $\ebar_T$.
\par
We present in this paper cross section limits for processes with an
energetic  jet plus large $\ebar_T$. This signature is characteristic of
processes not described by the  Standard Model, such as 
the  production of light 
gravitino pairs plus one jet \cite{brig}.
The data sample used for this analysis was
collected with the CDF detector from 1994 to 1995, and corresponds to a 
total integrated luminosity of 
$87$ pb$^{-1}$.
The CDF detector
is described in detail elsewhere \cite{CDF}; only features essential to
this analysis are summarized here.
The momenta of charged particles are measured 
in the
central tracking chamber (CTC), which is inside a $1.4$ T superconducting
solenoidal magnet. Outside the CTC, electromagnetic and hadronic 
calorimeters, which are segmented in $\eta-\phi$ towers and
 cover the pseudorapidity region $|\eta|<4.2$ \cite{eta}, 
are used to identify jets and electron candidates. Outside the
calorimeters, drift chambers in the region $\vert\eta\vert < 1.0$ provide muon
identification.

Events for this analysis passed a multilevel  trigger system which 
selected events with $\ebar_T\ge 35$ GeV.
$\ebar_T$ is defined to be the magnitude of the vector sum of transverse
energy in all calorimeter towers with $\vert\eta\vert\le 3.6$\cite{met}.
After removing cosmic ray and  accelerator related 
backgrounds\cite{smaria} we select events with $\ebar_T\ge 50$ GeV,  
at least one jet\cite{jet} with transverse energy $E_T\ge 10$ 
GeV in  the central region,  $\vert\eta\vert\le 0.7$, and
 with the additional request of $E_T\ge 80$ GeV 
for the most energetic jet. 
These requirements define the topology we are looking for and reduce
the presence of unphysical backgrounds.
\par
The backgrounds expected from Standard Model sources
are  due to
 $W+jet$ or $Z+jet$ processes plus a small contribution from $t\bar t$ and 
diboson ($WW,WZ,ZZ$) production. We estimate these with the 
PYTHIA\cite{pyt} generator 
 and a full simulation of the CDF detector. The 
 cross sections
for the $W/Z+jet$ processes 
are taken, for each value of jet multiplicity,
 from CDF measurements\cite{prl_w,prl_z}.
The cross sections for $t\bar t$ and diboson processes 
 are taken from theory\cite{tts,vvs}. 
The contribution from all these processes is reduced 
 by rejecting events 
containing electrons or muons with large transverse momentum, $P_T$: 
$P_T\ge 10$ GeV/c for electrons;
$P_T\ge 5$ GeV/c for muons 
or $P_T\ge 10$ GeV/c if the muon is within 
$\Delta R\equiv\sqrt{(\Delta\eta)^2+(\Delta\phi)^2}=1$ from a jet. Additional
 rejection is obtained by removing events 
which contain a jet with a ratio of electromagnetic to total energy larger
 than 0.95 or isolated tracks of $P_T\ge 30$ GeV/c. Here an isolated track is
defined as a track for
which the $\sum P_T$  of additional tracks
 within a cone of radius $\Delta R=0.4$ is smaller than 10 GeV/c. 
A total of $16,019$ events pass these requirements. 
\par

\begin{figure}[htb] \centering \mbox{}
\begin{minipage}[b]{2.95in} \centering 
\epsfxsize=2.95in \leavevmode
\epsfbox{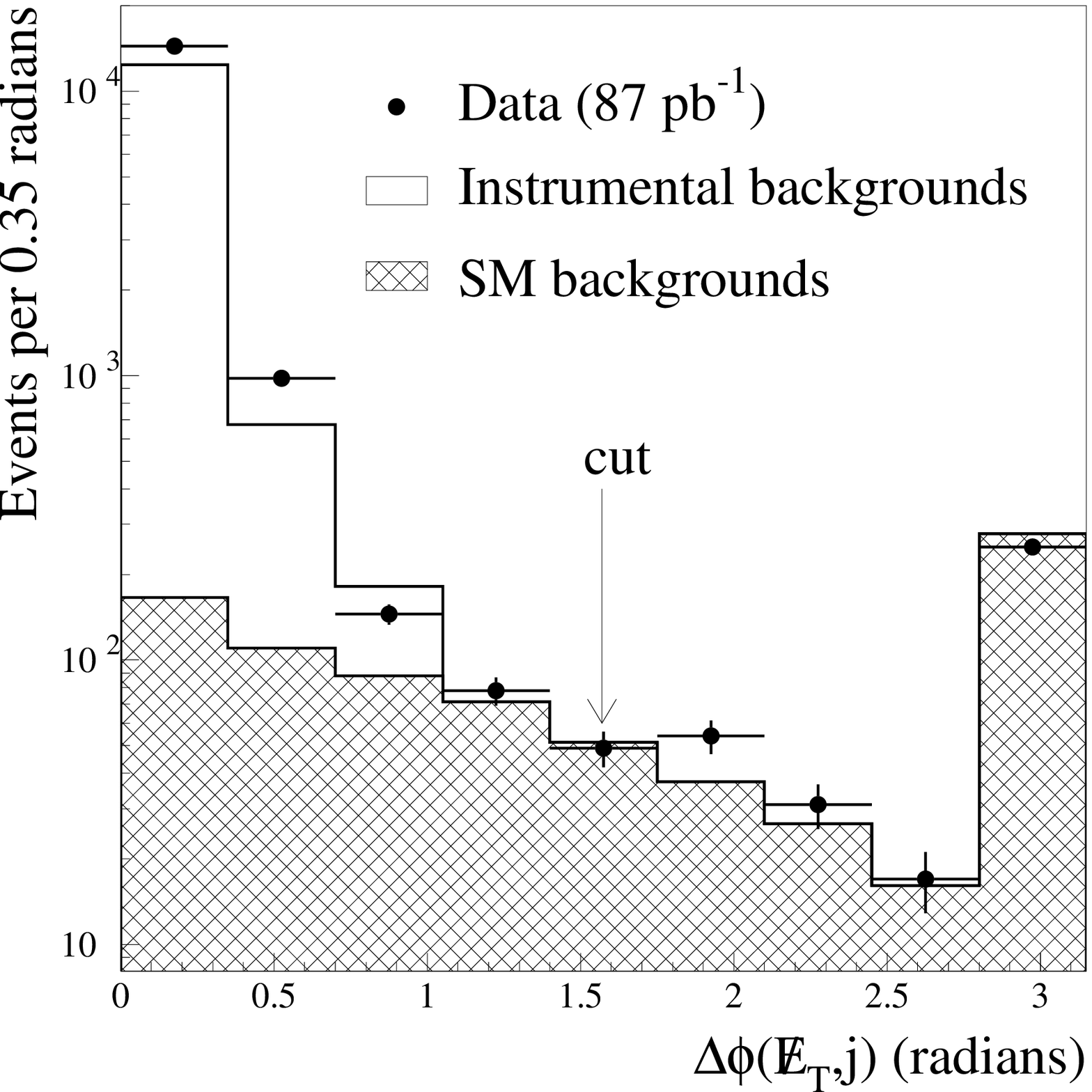}
\caption{The distribution of $\Delta\phi(\ebar_T,j)$
 for events before the
$\Delta\phi$ cut is applied (points), compared
to the instrumental backgrounds inferred from the control sample
 (white area) and the Standard Model
 backgrounds (shaded area).}
\label{dphi}
\end{minipage}
\end{figure}

The resulting data sample is dominated by instrumental backgrounds, 
 due to mismeasurement of otherwise balanced QCD multijet events.
The behavior  of these backgrounds
 is studied using a control sample passing a
trigger which selected 1/40 of the events
 having at least one jet with $E_T\ge 50$ GeV. Apart from  prescaling, 
the kinematical requirements we impose 
guarantee full overlap with the signal sample.
To reduce the instrumental backgrounds we cut on the azimuthal 
angle, $\Delta\phi (\ebar_T, j)$,
 between the direction of  $\ebar_T$ and the nearest jet.
 The distributions of $\Delta\phi (\ebar_T, j)$ for the data, the control 
sample,
 and for the Standard Model processes considered
are shown in Fig. \ref{dphi}. 
The requirement  
$\Delta\phi (\ebar_T, j)\ge 1.57$ radians  
is very effective in removing the
 instrumental backgrounds: from an extrapolation of the behaviour of the
control sample, and loosening the $\ebar_T$ cut to populate the tails of the
distribution, we derive that at most 14 events (at 95\% confidence level,
 C.L.) from instrumental
backgrounds are expected to survive this cut with respect to a total of 379
 events selected. The number of events expected from Standard Model 
processes is 
$380\pm 129$,  with contributions mainly from  $Z+jet$  ($204\pm 69$)  
and $W+jet$ 
($171\pm 57$) processes. 
Backgrounds from cosmic rays or beam halo have been considered and found 
negligible\cite{smaria} 
($\le 4$ events at 95\% C.L.). 
One source of uncertainty on the background estimate derives from 
uncertainties on  the production cross
sections ($20\%$ for $W/Z+jet$ cross sections\cite{prl_w,prl_z}, 
$20\%$ for $t\bar t$ corresponding
to the range of theoretical calculations\cite{tts} and $30\%$ for 
diboson production related to the use of different sets of parton
distribution functions\cite{vvs}). Another contribution comes from
the uncertainty on the selection 
acceptance (25\%) due mainly to the uncertainty on the jet energy scale
 (from 5\% for low-$E_T$ jets to 3\% for high-$E_T$ ones, reflecting the
uncertainty in our knowledge of the reconstructed jet energy\cite{jet}). 
A minor
contribution of 4\% comes from the uncertainty on
 the integrated luminosity. 
\par
\begin{figure}[htb] \centering \mbox{}
\begin{minipage}[b]{2.95in} \centering 
\epsfxsize=2.95in \leavevmode
\epsfbox{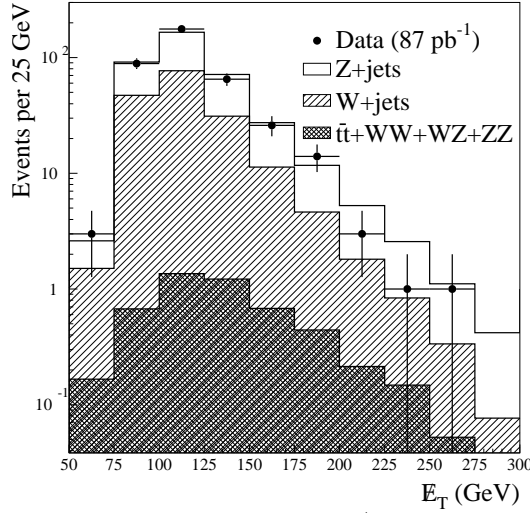}
\caption{The distribution of  $\ebar_T$ for data (points) 
compared to the central value 
expectations from $Z+jet$ events (white area), 
 $W+jet$ events (shaded area) and  for $t\bar t+WW+WZ+ZZ$
events (hatched area).} 
\label{met}
\end{minipage}
\end{figure}
\begin{figure}[htb] \centering \mbox{}
\begin{minipage}[b]{2.95in} \centering 
\epsfxsize=2.95in \leavevmode
\epsfbox{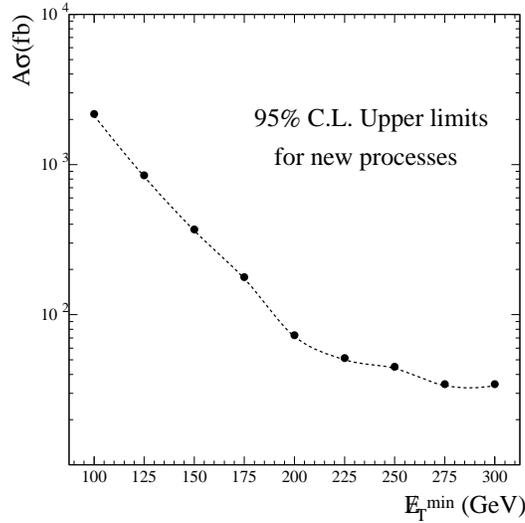}
\caption{The 95\% C.L. upper limits on the product of acceptance (A)
times  cross section  ($\sigma$) for the production of new 
processes.}
\label{met_lim}
\end{minipage}
\end{figure}

A jet correction algorithm\cite{corr} is applied to $\ebar_T$ 
 which takes into account
calorimeter nonlinearities and reduced response at boundaries
 between modules and calorimeter subsystems. 
 No correction is applied for high-$P_T$ muons
 since we remove the events containing them. 
Fig. \ref{met} shows the $\ebar_T$ distribution; the  data and  
the  expectation for Standard Model processes are in  good agreement.
The 95\% C.L.
 upper limits on the product of acceptance times   cross section 
for the production of physics beyond the Standard Model
are obtained using a Monte Carlo technique \cite{gluinos} which
convolutes the uncertainties on the integrated luminosity with
 background 
expectations. The limits, as a function of the $\ebar_T$ threshold, 
$\ebar_T^{min}$, are shown
in Fig.  \ref{met_lim}.
Systematic uncertainties on the acceptance are not included because they
depend on the particular physics process under consideration.
\par
The production of gravitinos, $p\bar p\to {\tilde G}{\tilde G}g, 
~ {\tilde G}{\tilde G}q$, is  simulated in  
HERWIG\cite{herw} by including the calculated matrix
 elements\cite{brig}, followed by a detector simulation. For the generation
the following inputs are used: 
$\sqrt{F}=200$ GeV; factorization/renormalization scale, $\mu$, equal to
the transverse energy of the emitted quark/gluon; and the MRSD$^\prime$
 set of parton
distribution functions\cite{mrs}. For such a choice of parameters,  
the production cross section, evaluated for 
$P_T^{{\tilde G}{\tilde G}}\ge 100$ GeV/c, 
amounts to $12.6\pm 4.0$ pb, where $P_T^{{\tilde G}{\tilde G}}$ 
is the transverse magnitude of the vector sum of the two gravitino momenta
before any further radiation has occurred.
The uncertainty on the cross section 
 has several contributions
which are added in quadrature:
 (i) 30\%  due
to  the choice of  factorization/renormalization scale ($\mu=2 E_T$ vs
$\mu=E_T/2$); 
(ii) 10\% due to the gluon 
radiation modeling in the Monte Carlo, obtained by comparing the
cross sections 
 before and after
radiation occurs;
(iii) and 5\% due to the choice of
 parton distribution function
 (e.g. MRSD$^\prime$ vs CTEQ2M\cite{cteq}).
The signal acceptance  is
the fraction of events with $P_T^{{\tilde G}{\tilde G}}\ge 100$ GeV/c 
which pass the  selection with a cut on $\ebar_T$. 
\par
\begin{figure}[htb] \centering \mbox{}
\begin{minipage}[b]{2.95in} \centering 
\epsfxsize=2.95in \leavevmode
\epsfbox{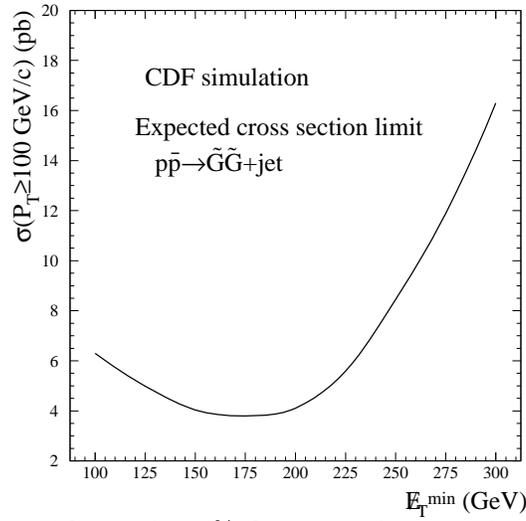}
\caption{The 95\% C.L. upper limit on the cross section 
for ${\tilde G}{\tilde G}+jet$ events with $P_T^{{\tilde G}{\tilde
G}}\ge 100$ GeV/c expected for no signal.}
\label{exp_xsec}
\end{minipage}
\end{figure}

We use the Monte Carlo technique mentioned above and convolute the uncertainty
on the acceptance with background estimates to derive  the
upper limit on the production cross section for ${\tilde G}{\tilde
G}+jet$ events with $P_T^{{\tilde G}{\tilde G}}\ge 100$ GeV/c
(see Fig. \ref{exp_xsec}). 
The best sensitivity (i.e. the smallest
upper limit on the  cross section) is reached for $\ebar_T^{min}= 175$ GeV. 
For such a threshold,
the acceptance amounts to $(6.2\pm 1.2)\%$, where the uncertainty is dominated
by the choice of absolute energy scale and the modeling of initial or
 final state gluon
radiation. Apart from the $\ebar_T$ threshold, the other selection 
criteria have a 
relative efficiency of about 80\%, 
essentially due to the requirement of a central jet and to the 
$\Delta\phi (\ebar_T, j)$ cut.
With 19 events selected above the optimized 
$\ebar_T$ threshold  of 175 GeV and an estimated background of 
$21.6\pm 7.0$ events, we derive the  95\% C.L. upper limit on the  
 signal of $16.9$ events, accounting for the 20\% relative uncertainty
 on the acceptance. This signal corresponds to an upper limit 
on the production cross section of 
3.1 pb for $P_T^{{\tilde G}{\tilde G}}\ge 100$ GeV/c.
In an ensemble of pseudo-experiments with the expected background and no true 
signal we would obtain this limit or better 35\% of the time.
Considering the $1/F^4$ dependence of the production cross section we derive,
from the comparison to the theory, 
a 95\% C.L. limit 
$\sqrt{F}\ge 217$ GeV. 
 Such a limit corresponds to a gravitino mass larger than  
$1.1\times 10^{-5}$ eV/c$^2$. We note that these limits are independent
of any unmeasured parameters; they would become stronger
if other supersymmetric particles were produced. 
\par
In conclusion, we have compared  events
containing large $\ebar_T$ and at least one energetic jet  
to the expectations from Standard Model processes and instrumental 
backgrounds. 
 The $\ebar_T$ distribution has been used to derive upper limits 
on the product of acceptance times 
cross section for the production of new processes beyond the Standard Model. 
We have selected 19 events with $\ebar_T\ge 175$
GeV with respect to an estimated background of 
$21.6\pm 7.0$  events. This
implies a 95\% C.L. upper limit on the
cross section of 3.1 pb for the production of ${\tilde G}{\tilde G}+jet$
events with $P_T^{{\tilde G}{\tilde G}}\ge 100$ GeV/c.
Comparing this number to the theoretical cross section   
we have derived the 95\% C.L.
 limit $\sqrt{F}\ge 217$ GeV, which corresponds to a gravitino mass 
$m_{\tilde G}\ge 1.1\times 10^{-5}$ eV/c$^2$. This limit 
is comparable to  
LEP measurements\cite{lep} based on events with photons and 
missing energy.
\par
     We thank the Fermilab staff and the technical staff of the
participating institutions for their vital contributions.  This work was
supported by the U.S. Department of Energy and National Science Foundation;
the Italian Istituto Nazionale di Fisica Nucleare; the Ministry of Education,
Science, Sports and Culture of Japan; the Natural Sciences and Engineering
Research Council of Canada; the National Science Council of the Republic of
China; the Swiss National Science Foundation; the A. P. Sloan Foundation; the
Bundesministerium fuer Bildung und Forschung, Germany; and the Korea Science
and Engineering Foundation.


\par

\begin{references}
\bibitem{FF} $F=\sqrt{3}m_{\tilde G}\bar M_P$, where
$\bar M_P=(8\pi G_N)^{-\frac{1}{2}}=2.4\times 10^{18}$ GeV/$c^2$
is the reduced Planck mass.
\bibitem{eta}
In the CDF coordinate system, the $z$ axis lies along the beam line.
The pseudorapidity $\eta$ is defined as $-\ln\tan(\theta/2)$, 
where $\theta$ is the polar angle with respect to the
proton beam direction.  The azimuthal angle with respect to the $z$ axis
is denoted by $\phi$.
\end{references}
\end{document}